\documentclass[aps,prb,reprint,preprintnumbers,amsmath,amssymb,showpacs,superscriptaddress,citeautoscript,floatfix]{revtex4-1}
\usepackage{amssymb}

\usepackage{graphicx}% Needed to include png/jpg/pdf figure files
\usepackage{float}
\usepackage{amsmath}
\usepackage{bm}
\usepackage{color, soul}
\usepackage[normalem]{ulem}
\usepackage{siunitx}
\usepackage{upgreek}
\usepackage[usenames,dvipsnames]{xcolor}

\usepackage[breaklinks]{hyperref}%hyperef package does internal and external linking (labels, urls)
\usepackage[all]{hypcap}%addition to hyperref, it ensures figure is correctly displayed (and not only the caption) when link is clicked
\hypersetup{
    plainpages=false,
    unicode=false,          % non-Latin characters in AcrobatĂŻÂżÂ˝s bookmarks
    pdfmenubar=true,        % show AcrobatĂŻÂżÂ˝s menu?https://www.overleaf.com/13403665mnpkhvmzmbnjhttps://www.overleaf.com/13403665mnpkhvmzmbnj
    pdffitwindow=false,     % window fit to page when opened
    pdfstartview={FitH},    % fits the width of the page to the window
    pdftitle={MoSe2_Quadrupoles},    % title
    pdfauthor={Alessandro Surrente},     % author
    pdfproducer={PWr}, % producer of the document
    pdfkeywords={High} {Magnetic} {Fields}, % list of keywords
    pdfnewwindow=true,      % links in new window
    linktoc=section,
    colorlinks=true,       % false: boxed links; true: colored links
    linkcolor=blue,          % color of internal links
    citecolor=red,        % color of links to bibliography
    filecolor=magenta,      % color of file links
    urlcolor=blue           % color of external links
}

\DeclareUnicodeCharacter{2212}{-}

\begin{document}

\title{Quadrupolar Excitons in \texorpdfstring{MoSe$_2$}{MoSe2} Bilayers}

\author{Jakub Jasi{\'n}ski}
\affiliation{Department of Experimental Physics, Faculty of Fundamental Problems of Technology, Wroclaw University of Science and Technology, 50-370 Wroclaw, Poland}
\affiliation{Laboratoire National des Champs Magn\'etiques Intenses, EMFL, CNRS UPR 3228, Universit{\'e} Grenoble Alpes, Universit{\'e} Toulouse, Universit{\'e} Toulouse 3, INSA-T, Grenoble and Toulouse, France}

\author{Joakim Hagel}
\affiliation{Department of Physics, Chalmers University of Technology, 412 96 Gothenburg, Sweden\\}

\author{Samuel Brem}
\affiliation{Department of Physics, Philipps-Universit{\"a}t Marburg, Renthof 7 35032 Marburg, Germany}

\author{Edith Wietek}
\affiliation{Institute of Applied Physics and Würzburg-Dresden Cluster of Excellence ct.qmat,
Technische Universität Dresden, 01062 Dresden, Germany}

\author{Takashi Taniguchi}
\affiliation{Research Center for Materials Nanoarchitectonics, National Institute for Materials Science, 1-1 Namiki, Tsukuba 305-0044, Japan}

\author{Kenji Watanabe}
\affiliation{Research Center for Electronic and Optical Materials, National Institute for Materials Science, 1-1 Namiki, Tsukuba 305-0044, Japan}

\author{Alexey Chernikov}
\affiliation{Institute of Applied Physics and Würzburg-Dresden Cluster of Excellence ct.qmat,
Technische Universität Dresden, 01062 Dresden, Germany}

\author{Nicolas Bruyant}
\affiliation{Laboratoire National des Champs Magn\'etiques Intenses, EMFL, CNRS UPR 3228, Universit{\'e} Grenoble Alpes, Universit{\'e} Toulouse, Universit{\'e} Toulouse 3, INSA-T, Grenoble and Toulouse, France}

\author{Mateusz Dyksik}
\affiliation{Department of Experimental Physics, Faculty of Fundamental Problems of Technology, Wroclaw University of Science and Technology, 50-370 Wroclaw, Poland}

\author{Alessandro Surrente}
\affiliation{Department of Experimental Physics, Faculty of Fundamental Problems of Technology, Wroclaw University of Science and Technology, 50-370 Wroclaw, Poland}

\author{Micha{\l} Baranowski}
\affiliation{Department of Experimental Physics, Faculty of Fundamental Problems of Technology, Wroclaw University of Science and Technology, 50-370 Wroclaw, Poland}

\author{Duncan K.\ Maude}
\affiliation{Laboratoire National des Champs Magn\'etiques Intenses, EMFL, CNRS UPR 3228, Universit{\'e} Grenoble Alpes, Universit{\'e} Toulouse, Universit{\'e} Toulouse 3, INSA-T, Grenoble and Toulouse, France}

\author{Ermin Malic}
\affiliation{Department of Physics, Philipps-Universit{\"a}t Marburg, Renthof 7 35032 Marburg, Germany}

\author{Paulina Plochocka}\email{paulina.plochocka@lncmi.cnrs.fr}
\affiliation{Department of Experimental Physics, Faculty of Fundamental Problems of Technology, Wroclaw University of Science and Technology, 50-370 Wroclaw, Poland}
\affiliation{Laboratoire National des Champs Magn\'etiques Intenses, EMFL, CNRS UPR 3228, Universit{\'e} Grenoble Alpes, Universit{\'e} Toulouse, Universit{\'e} Toulouse 3, INSA-T, Grenoble and Toulouse, France}

\date{\today}

\begin{abstract}
\href{paulina.plochocka@lncmi.cnrs.fr}{*paulina.plochocka@lncmi.cnrs.fr}\\
 \\
The quest for platforms to generate and control exotic excitonic states has greatly benefited from the advent of transition metal dichalcogenide (TMD) monolayers and their heterostructures. Among the unconventional excitonic states, quadrupolar excitons -- a superposition of two dipolar excitons with anti-aligned dipole moments -- are of great interest for applications in quantum simulations and for the investigation of many-body physics. Here, we unambiguously demonstrate the emergence of quadrupolar excitons in natural MoSe$_2$ homobilayers, whose energy shifts quadratically in electric field. In contrast to trilayer systems, MoSe$_2$ homobilayers
have many advantages, which include a larger coupling between dipolar excitons. Our experimental observations are complemented by many-particle theory calculations offering microscopic insights in the formation of quadrupolar excitons. Our results suggest TMD homobilayers as ideal  platform for the engineering of excitonic states and their interaction with light and thus candidate for carrying out on-chip quantum simulations.

\end{abstract}

\maketitle

\section*{Introduction}

Two-dimensional (2D) layered semiconductors, such as transition metal dichalcogenides (TMDs), have emerged as an ideal playground to study exciton physics on the nanoscale, essentially due to the intricate valley physics and the greatly enhanced electron-hole attraction related to the reduced dimensionality and dielectric screening  in the monolayer limit \cite{chernikov2014exciton,he2014tightly,hanbicki2015measurement,wang2018colloquium,mueller2018exciton, PhysRevLett.108.196802,perea2022exciton}. The subsequent development of van der Waals heterostructures significantly enriched this field of research. The absence of the lattice-matching constraints for TMDs and many other emerging layered 2D materials opens a new paradigm in material engineering, where different materials can be seamlessly stacked into virtually limitless combinations, 
whose properties can be tuned by both the material selection and the relative orientation \cite{geim2013van, novoselov20162d,ciarrocchi2022excitonic,huang2022excitons}.

For instance, homobilayers and heterostructures support long-lived dipolar interlayer excitons (IXs), where electrons and holes reside in different layers \cite{rivera2015observation,rivera2018interlayer,lee2014atomically,gerber2019inter,jiang2021interlayer,arora2017interlayer,merkl2019ultrafast}, and hence can be can be easily tuned by external electric field \cite{wang2018electrical,peimyoo2021electrical,liu2020electrically,tagarelli2023electrical, tagarelli2023electrical}. Transition metal dichalcogenide heterostructures have emerged as an excellent solid-state platform for exploring many-body physics and quantum phases arising from monopolar\cite{smolenski2021signatures} and dipolar interactions \cite{eisenstein2004bose, wang2019evidence,fogler2014high, regan2020mott,tang2020simulation,huang2021correlated,gu2022dipolar,zhou2021bilayer,xu2020correlated, brem2024optical}, entering fields traditionally dominated by ultracold atoms \cite{anderson1995observation, bloch2008many, bloch2012quantum, natale2019excitation, tanzi2019observation}.
Very recently it has been demonstrated that TMD heterostructures can also host more complex quasiparticles, referred to as quadrupolar excitons \cite{yu2023observation,li2023quadrupolar,lian2023quadrupolar, bai2023evidence,xie2023bright, slobodkin2020quantum,astrakharchik2021quantum,yu2023observation,li2023quadrupolar, deilmann2024quadrupolar}, stemming from the hybridization between two dipolar excitons with opposite dipole moments. The higher-order symmetry causes the quadrupole-quadrupole interactions to be substantially different compared to the dipole-dipole ones.
In particular, their non-local interactions can be finely tuned by the application of electric field. The quadrupolar interactions enable new collective phenomena beyond monopolar and dipolar interactions such as the exotic rotons, new flavors of Bose-Einstein condensate, charge density wave or topological superfluids\cite{weber2003bose,lahrz2014detecting,lahrz2015exotic,bhongale2013quantum,slobodkin2020quantum}. The solid-state matrix endows the quadrupolar states with robustness, which instead eludes their molecular counterparts\cite{carr2009cold,moses2017new}, and is particularly attractive to implement quantum simulation protocols and to reveal unconventional quantum states, many body phases\cite{bhongale2013quantum,lahrz2014detecting,lahrz2015exotic}, and phase transitions\cite{slobodkin2020quantum,astrakharchik2021quantum}.
The multipolar character of excitons in TMD structures can be continuously tuned between quadrupolar and dipolar states due to the nonlinear Stark effect\cite{li2023quadrupolar,yu2023observation} or mixing with other excitonic species (as we show herein), which  enables a continuous control over many-body interactions.

So far the formation of quadrupolar states in TMD systems has been explored (both theoretically and experimentally) only for TMD heterotrilayers \cite{yu2023observation,li2023quadrupolar,lian2023quadrupolar, bai2023evidence,xie2023bright, slobodkin2020quantum,astrakharchik2021quantum,yu2023observation,li2023quadrupolar, deilmann2024quadrupolar}, which enforce the formation of interlayer excitons with anti-aligned static dipole moments. Similar conditions can be found in TMD homobilayers \cite{arora2017interlayer,gerber2019inter,hagel2022electrical,feng2024highly}, suggesting that quadrupolar excitons might also form in these structures. However, they have remained elusive so far.

Here, we demonstrate the existence of quadrupolar states in a natural 2H-stacked homobilayer of MoSe$_2$. In our double gated device we identify two types of interlayer transitions with dipolar and quadrupolar character. Combining many-particle theoretical modelling, and electric field dependent reflectivity measurements, we provide a microscopic understanding of the complex excitonic landscape in a natural MoSe$_2$ bilayer. We show that the quadrupolar states emerge from the coupling between the dipolar transitions. The observed excitonic states, including the quadrupolar excitons, can be effectively tuned, with the use of electric field, between interlayer, hybrid and intralayer character showing that natural MoSe$_2$ bilayers are promising candidates to study many-body physics driven by field-tunable electric multipolar interactions.  
%=====================================================

%Figure1 sample design
\begin{figure*}[t]
\centering
\includegraphics[width=1.0\linewidth]{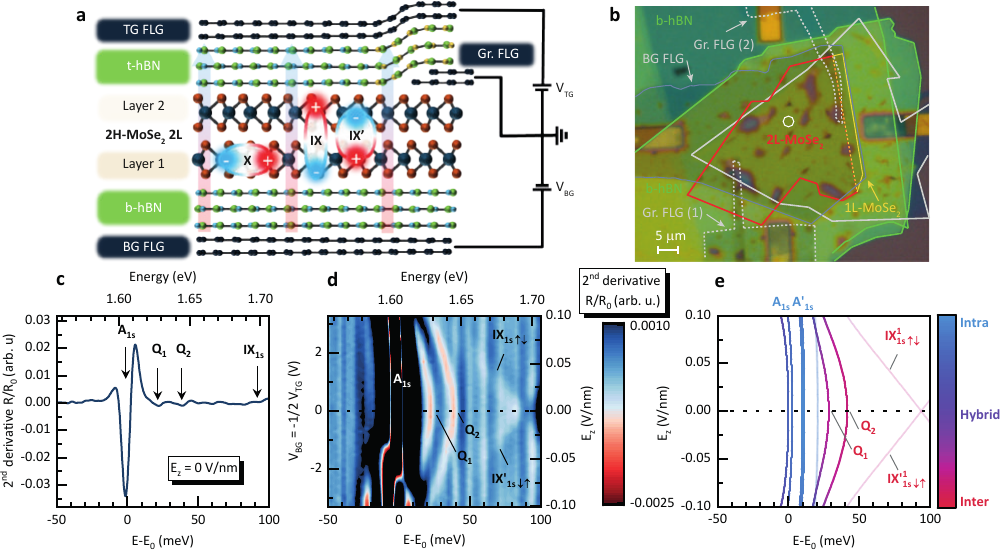}
\caption{\textbf{Natural MoSe$_2$ bilayer device and optical response under applied out-of-plane electric field.} \textbf{a} Schematic of the device. 2H-stacked natural bilayer of MoSe$_2$ encapsulated by insulating bottom and top hBN layers. Few layer graphene (FLG)  serve as the bottom (BG) and top (TG) gates, respectively, forming a capacitor-like structure. Two additional FLG flakes physically touch the MoSe$_2$ bilayer for grounding purposes (Gr.\ FLG 1/2). The effect of an applied out of plane electric field ($E_z$) indicated by the arrows) is shown on the intralayer exciton (X), and the interlayer excitons with opposite dipole moments (IX and IX').
\textbf{b} Micrograph of the sample. The white circle indicates the measurement spot. \textbf{c} Second derivative of normalised reflectivity ($R/R_0$) at $E_z=0$\,V/nm. \textbf{d} False-color map of the 2$^{\text{nd}}$ derivative of reflectivity as a function of the gate voltages (left axis) and the corresponding $E_z$ (right axis). The intensity of the strongest neutral A$_\textrm{1s}$ exciton transition is intentionally saturated to reveal the behaviour of the weaker interlayer transitions. Note, that at negative electric fields, $E_z<-0.5$\,V/nm, the device is unintentionally doped, thus for simplicity, we focus on the analysis of the positive $E_z>0$ electric field data. \textbf{e} Calculated evolution of the excitonic energy landscape under the influence of the electric field. The color scale corresponds to the spatial character of the excitons i.e. intralayer (blue), hybrid (purple) or interlayer (red). The opaque (semi-transparent) lines correspond to the spin-singlet (spin-triplet) states. The bottom energy scales ($E-E_0$) in panels \textbf{c} and \textbf{d} are shown with respect to the A$_{1\text{s}}$ exciton energy $E_0=1.606$\,eV.}
\label{fig:Design}
\end{figure*} 

\section*{Results and Discussion}
\subsection*{Observation of quadrupolar excitons}
We have investigated a natural 2H-stacked MoSe$_2$ homobilayer, fully encapsulated in hexagonal boron nitride (hBN). The hBN encapsulated  MoSe$_2$ is grounded by a few layer graphite (FLG) electrodes. Two additional FLG electrodes are used as the top and the bottom gates. Schematic and microscope images of the device are shown in  Fig.\,\ref{fig:Design}(a-b). 
During the fabrication of the device by the dry-transfer method, the stack was annealed after each stamping step. The goal of annealing was to minimize the concentration of bubbles and simultaneously improve the adhesion between the consecutive layers (see Methods section for more details on the fabrication procedure).
To reveal the complex excitonic landscape of the natural MoSe$_2$ bilayer, we studied its optical response as a function of the out-of-plane electric field ($E_z$) using the capacitor-like design of the structure which allows for the independent control of the out-of-plane electric field and carrier doping (see Methods for details).

A typical reflectivity spectrum of bilayer MoSe$_2$ (measured at temperature of 5\,K), shown as a second derivative is presented in Fig.\,\ref{fig:Design}(c). The spectrum is dominated by a strong resonance related to the A$_\textrm{1s}$ exciton state, accompanied on the high energy side by three weaker transitions, labelled as Q$_1$, Q$_2$ and IX$_\textrm{1s}$. To understand the origin of these transitions, we track their evolution as a function of the electric field. 
In Fig.\,\ref{fig:Design}(d) we present the reflectivity spectrum at varying electric field, plotted in the form of a false-color map.
Characteristic features can be identified in the false colour map, providing deeper insight into the exciton landscape and the mutual interaction of the excitonic states. The states labelled as IX$_{1\text{s},\uparrow\downarrow}$ and IX'$_{1\text{s},\downarrow\uparrow}$ exhibit a linear Stark shift, consistent with their dipolar, interlayer character \cite{kovalchuk2023interlayer,sung2020broken,shimazaki2020strongly, feng2024highly}. 
Matching the observed shift to the Stark shift simulated with the model detailed below, we estimate the dipole moment to be $d\simeq0.5-0.6$\,e$\cdot$nm, which is in the range of the dipole length reported for other MoSe$_2$ bilayers \cite{sung2020broken,kovalchuk2023interlayer}.
In addition, the new states labelled Q$_1$ and Q$_2$ exhibit a distinct, quadratic Stark shift at low electric fields. 
This behaviour is the unequivocal evidence of their quadrupolar nature \cite{yu2023observation,li2023quadrupolar, xie2023bright}, which stems from the coupling of a pair of anti-aligned dipolar states. The symmetric arrangement of charges in an electric quadrupole yields a zero dipole moment at $E_z=0$. However, increasing electric field displaces the charges, and the quadrupolar state gradually acquires a dipole moment, giving rise to the non-linear Stark shift. 
To corroborate the assignment of the Q$_1$ and Q$_2$ as quadrupolar excitons, we also plotted in Fig. S1 the electric field dependence of the static electric dipole moment, calculated as $\frac{\text{d}E}{\text{d}E_z}$. The non-linear dependence of the quadrupolar exciton energy on the electric field translates to a vanishingly small electric dipole at low fields. The static dipole moment increases with increasing electric field and steadily approaches the dipole moment obtained for the dipolar spin-triplet interlayer exciton IX$^{\text{A}}_{{1\text{s}}\uparrow\downarrow}$. At higher fields, around \SI{0.1}{\volt /\nano\metre}, one can observe the deviation from the expected behaviour which stems from the interaction with other excitonic states as we discuss in the further part of the manuscript.
Additionally, analogous excitonic resonances shifting quadratically with the applied electric field, which attests to their quadrupolar nature, were observed on different spots on the main device (Fig.\,S2) and also on a 2$^{\text{nd}}$ device shown in Fig.\,S3.

\begin{figure*}[t]
\centering
\includegraphics[width=1.0\linewidth]{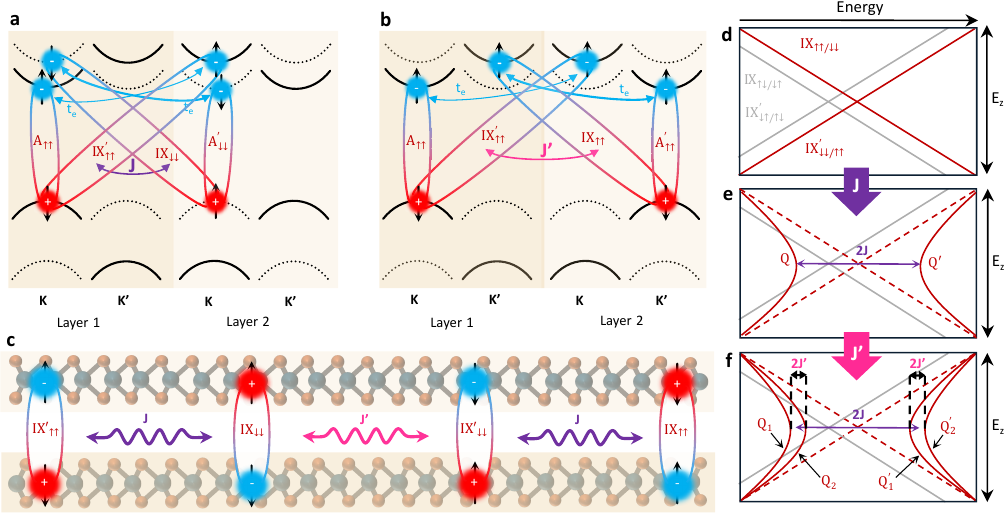}
\caption{\textbf{Dipolar exchange interaction in natural MoSe$_2$ bilayer.} \textbf{a} Schematic of the $J$ coupling between one of the oppositely aligned pairs of singlet IXs (same valley, different spin) and the interplay with electron tunneling to their respective A excitons. \textbf{b} Schematic of the $J'$ coupling between one of the oppositely aligned pairs of singlet IXs (different valley, same spin) and the interplay with electron tunneling to their respective A excitons. \textbf{c} Schematic of the real space interaction of the four degenerate singlet IXs via $J$ and $J'$ couplings. \textbf{d} Scheme of IX exciton species energy as a function of the electric field for the singlet ($\uparrow\uparrow,\downarrow\downarrow$) and triplet ($\uparrow\downarrow,\downarrow\uparrow$) IX species without $\tilde{J}$ coupling. \textbf{e} including the J coupling term IX singlet states mix forming quadrupole branches Q and Q$^\prime$.
(f) exciton landscape including both $J$ and $J'$ coupling which splits degeneracy of Q and Q$^\prime$ branches. The dashed lines in \textbf{e} and \textbf{f} correspond to spin-singlet states in the non-interacting picture \textbf{d}, from which the quadrupolar branches formed.}
\label{fig:coupling}
\end{figure*} 

\subsection*{Microscopic model}
To provide a detailed microscopic understanding of our observations, we complement our experiments with an effective many-particle model that allows for the identification of the key coupling mechanisms. The symmetric band structure in naturally stacked bilayers hosts a fourfold degeneracy, stemming from the combination of valley and layer degeneracy. This applies to all intralayer and interlayer exciton species, including the A-exciton, interlayer spin-singlet ($\uparrow\uparrow, \downarrow\downarrow$) and spin-triplet states ($\uparrow\downarrow, \downarrow\uparrow$) (see Supplementary Fig.\,S7(a,b,c)). For instance, the spin-singlet interlayer excitons (Supplementary Fig.\,S7(b)), IX$_{\uparrow\uparrow}$ and IX$_{\downarrow\downarrow}$ have a reversed dipole moment, i.e., exchanged positions of the electron and hole, with respect to the other two (IX'$_{\uparrow\uparrow}$ and IX'$_{\downarrow\downarrow}$), leading to their mixing via the dipole exchange interaction (see Section II in SI for further details). 
The effective Hamiltonian, which includes all possible interaction channels, can then be written as   
\begin{equation}
H=H_{0}+H_T+H_{QC}.\nonumber
\end{equation}
Here $H_0$ describes the electron and hole Coulomb interaction through
the generalized Wannier equation, \cite{ovesen2019interlayer} together with the exciton response to the external electric field. $H_T$ is the tunneling contribution, which takes into account both electron and hole tunneling \cite{hagel2021exciton}. The last term $H_{QC}$ contains the effective dipole exchange coupling $\tilde{J}$ (see Eq.\ (S5) in SI), giving rise to the formation of quadrupolar excitons.

We initially focus on the exchange coupling to explain the nonlinear shift of the Q$_1$ and Q$_2$ transitions.
We assume that $\tilde{J}$ only mixes the 1s interlayer exciton states of the spin-singlet configuration IXs ($\uparrow\uparrow,\downarrow\downarrow$) and opposite dipole moments. This tentative assumption is motivated by the more pronounced signature of Q-states compared to IX features, which suggests a higher oscillator strength characteristic for singlet transition. Moreover, the mixing is assumed to stem from the Coulomb interaction, which is a spin-conserving interaction. Such mixing
between the necessary spin-triplet states would not be
spin-conserving in a bilayer system. Nevertheless, qualitatively similar quadrupole formation could be expected assuming an equally efficient coupling between interlayer triplet states. We infer that four spin-singlet IXs mix through two possible interaction paths ($\tilde{J}=J+J'$) (See Eq.\ (8) in SI) schematically shown in Fig.\,\ref{fig:coupling}(a)-(c). The first path couples anti-aligned dipolar IXs corresponding to the same valley but with opposite spin configuration ($\text{IX}_{\uparrow\uparrow}+\text{IX}'_{\downarrow\downarrow}$ and $\text{IX}_{\downarrow\downarrow}+\text{IX}'_{\uparrow\uparrow}$), schematically drawn in Fig.\,\ref{fig:coupling}(a,c) and denoted as $J$. The second path, indicated as $J'$, mixes anti-aligned dipolar IXs localized in the opposite valleys, but with the same spin configuration ($\text{IX}_{\uparrow\uparrow}+\text{IX}'_{\uparrow\uparrow}$ and $\text{IX}_{\downarrow\downarrow}+\text{IX}'_{\downarrow\downarrow}$), as schematically represented in Fig.\,\ref{fig:coupling}(b,c). 

The evolution of the energy landscape of IXs under electric field in the absence and in the presence of the exchange couplings $J/J'$ is schematically presented in Fig.\,\ref{fig:coupling}(d-f). 
When the $J/J'$ couplings are not accounted for (Fig.\,\ref{fig:coupling}(d)), the application of an electric field
gives rise to two linearly shifting IX states: the higher energy spin-singlet and the lower energy spin-triplet states, offset by the spin-orbit coupling in the conduction band. The inclusion of the first term $J$ in Fig.\,\ref{fig:coupling}(e) mixes the spin-singlet IXs, which yields two quadrupolar excitons with opposite curvature, i.e., the (symmetric -- red-shifting) Q and (anti symmetric -- blue-shifting) Q' separated by an energy $2J$ at zero electric field (and by $\pm J$ from the IX singlet states in the non-interacting picture). The $J'$ coupling, added in Fig.\,\ref{fig:coupling}(f), leads to a further splitting of the quadrupolar branches into Q$_1$ and Q$_2$, and Q'$_1$ and Q'$_2$, each pair separated by $2J'$ (see the detailed model description in Section II in the SI). 
As the lower energy spin-triplet IXs (IX$_{\uparrow\downarrow}$, IX$_{\downarrow\uparrow}$, IX'$_{\uparrow\downarrow}$, IX'$_{\downarrow\uparrow}$) are not spin-conserving, they remain unaffected by the $J/J'$ coupling and thus they follow the standard linear Stark shift.

\begin{figure*}[t]
\centering
\includegraphics[width=1.0\linewidth]{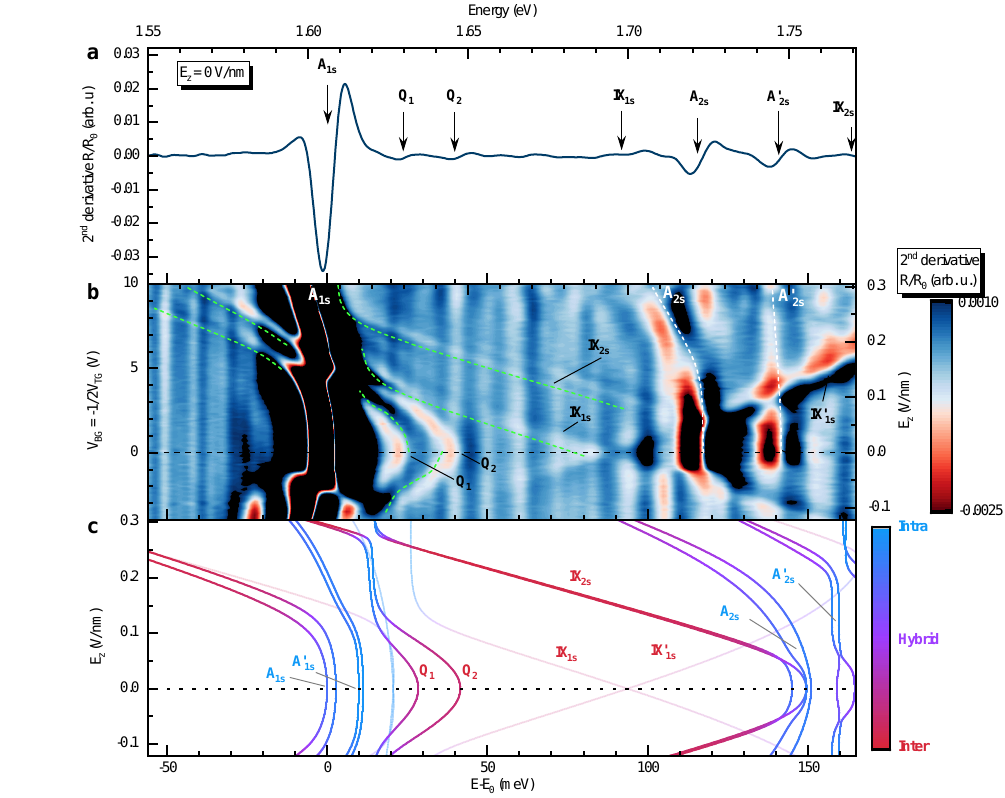}
\caption{\textbf{Extended exciton energy landscape under electric field.} 
\textbf{a} Second derivative of reflectivity without electric field. \textbf{b} False-color map of 2$^{\text{nd}}$ derivative of reflectivity as a function of applied gate voltages (left axis) and the corresponding $E_z$ (right axis). \textbf{c} Simulated exciton energy landscape in electric field which includes the hole and electron tunneling as well as the coupling $\tilde{J}$ that forms the quadrupoles. The color scale corresponds to the spatial character of the excitons i.e. intralayer (blue), hybrid (purple) or interlayer (red). The opaque (semi-transparent) lines correspond to the spin-singlet (spin-triplet) states. The bottom energy scales ($E-E_0$) in panels \textbf{a} and \textbf{b} are shown with respect to the A$_{1\text{s}}$ exciton energy $E_0=1.606$\,eV.}
\label{fig:quad}
\end{figure*}

By matching the values of $J$ (energy de-tuning from the spin-triplet IX plus spin orbit coupling of $\sim 22$\,meV \cite{kormanyos2015k} at $E_z=0$\,V/nm) and $J'$ (half of the separation between the Q$_1$ and Q$_2$ at $E_z=0$\,V/nm) to match the experimentally observed redshifting quadrupole branch, we obtain $J=90$\,meV and $J'=8$\,meV. The results of the simulation are shown in Fig.\,\ref{fig:Design}(e). The good qualitative agreement with the experiment summarized in Fig.\,\ref{fig:Design}(d) demonstrates that our effective Hamiltonian successfully explains the experimentally observed electric field-induced nonlinear energy shift of the Q$_1$ and Q$_2$ states.
These states, formed from linear combination of spin-singlet IXs (Q$_1 \sim \text{IX}'_{\uparrow\uparrow} + \text{IX}_{\downarrow\downarrow}$ and Q$_2 \sim \text{IX}'_{\downarrow\downarrow} + \text{IX}_{\uparrow\uparrow}$),
correspond to symmetric quadrupole branches which red shift with increasing electric field.
According to the presented analysis the IX states (IX$_{\uparrow\downarrow}$/IX$'_{\downarrow\uparrow}$) with resonance around 90\,meV above the A$_\textrm{1s}$ at $E_z = \SI{0}{\volt / \nano\metre}$ are transitions originating from optically bright spin-triplet states \cite{yu2018brightened}. Due to the lack of coupling $\tilde{J}$ these preserve their dipolar character exhibiting a linear Stark shift. Here we note that the opposite assignment of the dipolar transition origin can also be found\cite{feng2024highly}. Unfortunately, our model does not allow for a definitive differentiation between a singlet or triplet origin of quadrupolar states. Importantly the singlet or triplet nature of the interlayer is not essential to interpret the quadratic shift of $Q_1$ and $Q_2$ transitions.

\subsection*{Charge tunneling}
To reveal the importance of the charge tunneling mediated interaction between different excitonic species we plot the measured reflectivity spectra over an extended energy and electric field ranges as shown in Fig.\,\ref{fig:quad}(a-b). At higher electric field we observe an anti-crossing behaviour of A$_{1\text{s}}$ with interlayer exciton species such as the spin-triplet IX$_{1\text{s},\uparrow\downarrow}$, 2s spin-singlet IX$_{\text{2s},\uparrow\uparrow}$ and the quadrupolar Q$_{1/2}$ states as indicated by green dashed lines in Fig.\,\ref{fig:quad}(b).
To explain this behaviour we incorporate the electron tunneling term ($t_{\text{e}}$) into our model (schematically shown in Fig.\,\ref{fig:coupling}(a-b) by the arrows labeled $t_\text{e}$). 
In Fig.\,\ref{fig:quad}(c) we present the results of the simulation, taking the electron tunneling into account, which clearly shows that the anti-crossing behaviour is well captured by the model (see also Fig.\,S8 in SI showing progressively the contribution of the various coupling mechanisms to the exciton spectrum). Note that electron tunneling is usually considered symmetry-forbidden in naturally stacked homobilayers (H$^h_h$ stacking) \cite{wang2017interlayer}. Nevertheless, our results demonstrate that some electron tunneling occurs, and it is crucial for the correct description of the excitonic landscape in MoSe$_2$ bilayers under the electric field. In the simulation, we assume the electron tunneling term to be $t_{\text{e}}=11.9$\,meV, corresponding to the calculated value for the H$^X_h$ stacking \cite{hagel2021exciton}. 
The much stronger hole tunneling term ($t_{\text{h}}=56.3$\,meV \cite{hagel2021exciton}) predominantly drives the hybridization between various interlayer and intralayer states \cite{hagel2022electrical,sokolowski2023twist}. These manifest in the exchange of the oscillator strength with increasing electric field as they approach energetically (see also the extended energy range data in Supplementary Fig.\,S4(b-c), where the observed hole tunneling mediated hybridizations are marked). 

Another significant effect, stemming from the hybridization due to charge tunneling (both electron and hole), is reflected in the change of the intra-inter layer character of excitonic states, which is tuned by the value of electric field. This is shown in the simulated spectra of Fig.\,\ref{fig:quad}(c) as the color coding of the lines, where the blue, purple and red correspond to the intralayer, hybrid and interlayer character, respectively. For example, the electron tunneling changes the character of quadrupolar excitons when they approach the A$_\textrm{1s}$ transition with increasing electric field.  
Around the anti-crossing region, the quadrupoles rapidly change their character from interlayer to intralayer, with a negligible energy shift as a function of the electric field. At the same time, some of the A excitons acquire a partially interlayer character. This is in contrast to the heterotrilayer case, where the quadrupolar exciton is the lowest state of the system, and its energy shift steadily approaches a rate which is characteristic for the dipolar interlayer exciton \cite{yu2023observation, xie2023bright}. Similar electric field induced change of exciton spatial characters can be observed for other transitions (see also Fig.\,S4(c) in the Supplementary Information).   
The charge carrier tunneling also explains the suppression of the blue-shifting anti-symmetric quadrupole branch in our spectra as a result of mixing with A$_\textrm{2s}$ exciton states. 
The antisymmetric exciton branch is expected to be at energies very close to the A$_{\text{2s}}$ intralayer exciton at zero field. Due to this close proximity, the antisymmetric quadrupolar excitons hybridize with intralayer A$_{\text{2s}}$ exciton, and their signature in the optical spectrum vanishes. However, the mixing of these intra- and interlayer states manifests as a splitting of A$_{\text{2s}}$ exciton at zero electric field. This state is not expected to exhibit any splitting in the non-interacting picture, but in the presence of quadrupolar exciton complexes, it shows a finite splitting, following hybridization with interlayer excitons such as IX$_{\text{2s},\uparrow \uparrow}$ or the antisymmetric branch of the main quadrupolar excitons.
Consequently, the A$_\textrm{2s}$ state splits into A$_\textrm{2s}$ and A$'_\textrm{2s}$, even at zero electric field, as can be seen both in the experimental (Fig.\,\ref{fig:quad}(b)) and theoretical (Fig.\,\ref{fig:quad}(c)) spectra (see also Supplementary Fig.\,S8, where the influence of the individual couplings is shown). The conclusions drawn from the reflectivity spectra are supported by PL measurements shown in Fig.\,S5. In the evolution of the PL spectra in electric field, the characteristic anti-crossing behaviour when quadrupolar states approach the A$_\textrm{1s}$  exciton can be observed, together with the red shift of the A$_\textrm{2s}$ emission.  

\subsection*{Discussion}
The origin of the couplings $J/J'$ has so far been attributed to low density effects in the Coulomb Hamiltonian. As demonstrated, the exchange coupling does indeed lead to terms which mix different dipoles and qualitatively fits well within the picture that $J \gg J'$. This is since $J$ includes both long and short range electron-hole exchange, whereas the short-range interaction is symmetry forbidden in $J'$ due to the mixing of different valleys (see expression for $H_{QC}$ in Eq.\ (S8) in SI) \cite{yu2014dirac}. For intralayer excitons, this coupling has been calculated to be around 20\,meV \cite{qiu2015nonanalyticity} and is expected to be smaller for interlayer excitons due to the reduced wave function overlap between the layers. Other effects such as density dependent dipole-dipole attraction might play a role in enhancing the mixing between different dipoles. Taking into account the device-to-device variability of the properties of TMD-based devices, the clear experimental signatures of exciton quadrupole formation, together with the overall good qualitative agreement with theory, indicate that the coupling between the different dipoles are much stronger than previously thought.
Our effective Hamiltonian successfully explains the evolution of the exciton landscape in bilayer MoSe$_2$. The formation of a quadrupolar state is driven by the dipolar exchange interaction between interlayer spin-singlet states. At the same time, the interlayer triplet states preserve their dipolar character, shifting linearly in the electric field. In addition, the hole and electron tunneling are responsible for the observed avoided crossing behaviour, and hybridization of the states.    

In summary, we have investigated the evolution of the exciton energy landscape under external electric field in natural MoSe$_2$ homobilayers. Notably, for the first time, we observe quadrupolar exciton states in a natural MoSe$_2$ bilayer. These excitonic transitions, characterized by nonlinear shift in an electric field, exhibit a much stronger dipolar exchange interaction than the one observed in heterotrilayers. Our experimental observations are accurately captured by the proposed many-particle effective Hamiltonian. We propose that dipolar excitons are characterized by spin-triplet configuration, while quadrupolar states emerge from the exchange coupling of the interlayer spin-singlet excitons. Moreover, our model highlights the importance of hole and electron tunneling for understanding the exciton landscape evolution under the electric field. 

Our research underscores the potential of MoSe$_2$ bilayers to serve as a field-tunable exciton playground, wherein the mutual interaction of exciton states facilitates the effective tuning of their spatial and electric multipole characteristics via electric fields. Therefore, we show that natural MoSe$_2$ bilayers display potential to be considered as a solid-state platform to study many-body physics driven by field-tunable electric multipolar interactions. The inherent robustness of a homobilayer as compared to layer-by-layer stacking of TMD heterobi- or trilayers makes the platform proposed here easier to incorporate reliably into devices. This stems from the fact that homobilayers are not prone to imperfect flake alignment, flake rearrangement during deposition and possible post-stacking surface reconstruction for lattice commensurate stacks, which unavoidably plague other heterostructures.

\section*{Methods}
\subsection*{Sample fabrication}
The sample was fabricated using mechanically exfoliated flakes and stacked one by one using the dry transfer method.
Each step of layer deposition was followed by annealing in ambient conditions, by ramping the temperature from \SIrange{100}{150}{\celsius} for the duration of $\sim$15\,min. At the final step, the sample was annealed for $\sim$15\,min at \SI{200}{\celsius}. The goal of the annealing was to remove or coagulate air bubbles that notoriously form in TMD stacks during dry transfer deposition.
The MoSe$_2$ bilayer is encapsulated with hBN and sandwiched in between few layer graphite (FLG) layers acting as the bottom and top gates. Two additional FLG layers are connected directly to (physically touching) the MoSe$_2$ bilayer serving as the grounding contacts, one as a spare contact. All FLG contacts overlap the nearby evaporated gold paths through which the voltage is applied. The gold pads are pre-deposited on the SiO$_2$ substrate. Our sample design does not require any additional lithography after the stack is transferred on the substrate, which minimizes the risk of introducing defects and lower the optical quality of the sample.

\subsection*{Measurements}

The sample is wire bonded in a chip carrier installed in a custom-made electrical adapter for the cold finger inside a helium flow cryostat. All presented measurements were performed at cryogenic temperatures of $\sim \SI{5}{\kelvin}$.

We characterized the influence of the applied gate voltages at the bottom ($V_{BG}$) and top ($V_{TG}$) gates. To adjust for the unequal thicknesses of the bottom and top insulating layers of hBN, we found the optimal gate voltage ratio which minimizes the effect of free carrier doping during electric field sweep to be $ V_{BG} = - \frac{1}{2} V_{TG}$. To make sure that we keep the most neutral doping level we checked also the ratio of neutral to charged exciton (trion) by applying gate voltages of the same polarity ($V_{BG} = \frac{1}{2} V_{TG}$). We found the $V_{BG} = \frac{1}{2} V_{TG} = 0$\,V  to be the optimal initial voltages due to the highest ratio of neutral to charged exciton, both in PL and Reflectivity (Supplementary Fig.\,S6). The reflectivity measurements were performed using a Tungsten-Halogen white light source, while PL used a 532\,nm continuous wave laser at $\sim$\,1\,mW power.

\subsection*{Data analysis}
The details of the analysis of the reflectivity spectra are described in section III of the Supplementary Information. Fig.\,S9 shows the scheme of the data processing. The comparison of the reflectivity spectra in the form of $R/R_0$ and its 1$^{\text{st}}$ and 2$^{\text{nd}}$ derivatives are shown in Fig.\,S10.

The strength of the applied electric field was calculated by matching the dipole moments of the measured and simulated spin-singlet interlayer exciton for low electric field/gate voltages, far from the crossing region.

\subsection*{Theoretical model}
Exciton energies were modelled using an effective many-particle theory based on the density matrix formalism and input from density functional theory
\cite{shimazaki2020strongly}. A two-particle tunneling Hamiltonian is formulated and the excitonic response to the electric field is included to first order \cite{hagel2022electrical}. The exchange interaction giving rise to the quadropole formation is included from the low-density Coulomb Hamiltonian and matching the coupling strength to the experiment.

\section*{Data Availability}
The experimental and theoretical datasets generated and/or analysed during this study are available at \href{https://doi.org/10.5281/zenodo.14584540}{DOI:10.5281/zenodo.14584540}.

%from the corresponding author on reasonable request.

\bibliography{Bibliography}

\section*{Acknowledgements}
J.J.\ acknowledges funding from the National Science Centre Poland within the Preludium Bis 1 (2019/35/O/ST3/02162) program. M.B. acknowledges funding from the National Science Centre Poland within the Sonata Bis (2020/38/E/ST3/00194) program and OPUS LAP (2021/43/I/ST3/01357). All authors thank Marzia Cuccu and Sophia Terres for their support in the laboratory work. P.P acknowledge supported through the EUR grant NanoX no.\ ANR-17-EURE-0009 in the framework of the ``Programme des Investissements d’Avenir''. The Marburg  group (S.B. and E.M) acknowledges funding from the Deutsche Forschungsgemeinschaft (DFG) via SFB 1083 (project B9). K.W.\ and T.T.\ acknowledge support from the JSPS KAKENHI (Grant Numbers 21H05233 and 23H02052) and World Premier International Research Center Initiative (WPI), MEXT, Japan. A.C.\ and E.W.\ gratefully acknowledge funding from the Deutsche Forschungsgemeinschaft via SPP2244 grant (Project-ID: 443405595) and the Würzburg-Dresden Cluster of Excellence on Complexity and Topology in Quantum Matter (ct.qmat) (EXC 2147, Project-ID 390858490). E.M.\ and A.C.\ acknowledge DFG funding via project 542873285.

\section*{Author Contributions}
J. J. has carried out all optical experiments and drafted the text and figures of the main manuscript and the supplementary information. J. H. and S. B. under the supervision of E. M. developed the theoretical model and performed the simulations. E. W. and A. C. have provided the necessary training and participated in the fabrication of electrical devices and electrical measurements. T. T. and K. W. have provided the high-quality hexagonal boron nitride for encapsulation of the sample. N. B. has been involved in the optimization of the experimental setup for electrical measurements and involved in those measurements. M. D. A. S., D. K. M. contributed to data analysis, interpretation of results and manuscript preparation. M. B. and P. P. have proposed the goal of the scientific inquiry, the methodology and refined the manuscript text and the interpretation of the experimental results.

\section*{Competing Interests Statement}
Authors declare no competing interests.

\end{document}

\title{Supplementary Information: Quadrupolar Excitons in \texorpdfstring{MoSe$_2$}{MoSe2} Bilayers}

\author{Jakub Jasi{\'n}ski}
\affiliation{Department of Experimental Physics, Faculty of Fundamental Problems of Technology, Wroclaw University of Science and Technology, 50-370 Wroclaw, Poland}
\affiliation{Laboratoire National des Champs Magn\'etiques Intenses, EMFL, CNRS UPR 3228, Universit{\'e} Grenoble Alpes, Universit{\'e} Toulouse, Universit{\'e} Toulouse 3, INSA-T, Grenoble and Toulouse, France}

\author{Joakim Hagel}
\affiliation{Department of Physics, Chalmers University of Technology, 412 96 Gothenburg, Sweden\\}%

\author{Samuel Brem}
\affiliation{Department of Physics, Philipps-Universit{\"a}t Marburg, Renthof 7 35032 Marburg, Germany}

\author{Edith Wietek}
\affiliation{Institute of Applied Physics and Würzburg-Dresden Cluster of Excellence ct.qmat,
Technische Universität Dresden, 01062 Dresden, Germany}

\author{Takashi Taniguchi}
\affiliation{International Center for Materials Nanoarchitectonics, National Institute for Materials Science, Tsukuba, Ibaraki 305-004, Japan}

\author{Kenji Watanabe}
\affiliation{Research Center for Functional Materials, National Institute for Materials Science,
Tsukuba, Ibaraki 305-004, Japan}

\author{Alexey Chernikov}
\affiliation{Institute of Applied Physics and Würzburg-Dresden Cluster of Excellence ct.qmat,
Technische Universität Dresden, 01062 Dresden, Germany}

\author{Nicolas Bruyant}
\affiliation{Laboratoire National des Champs Magn\'etiques Intenses, EMFL, CNRS UPR 3228, Universit{\'e} Grenoble Alpes, Universit{\'e} Toulouse, Universit{\'e} Toulouse 3, INSA-T, Grenoble and Toulouse, France}

\author{Mateusz Dyksik}
\affiliation{Department of Experimental Physics, Faculty of Fundamental Problems of Technology, Wroclaw University of Science and Technology, 50-370 Wroclaw, Poland}

\author{Alessandro Surrente}
\affiliation{Department of Experimental Physics, Faculty of Fundamental Problems of Technology, Wroclaw University of Science and Technology, 50-370 Wroclaw, Poland}

\author{Micha{\l} Baranowski}
\affiliation{Department of Experimental Physics, Faculty of Fundamental Problems of Technology, Wroclaw University of Science and Technology, 50-370 Wroclaw, Poland}

\author{Duncan K.\ Maude}
\affiliation{Laboratoire National des Champs Magn\'etiques Intenses, EMFL, CNRS UPR 3228, Universit{\'e} Grenoble Alpes, Universit{\'e} Toulouse, Universit{\'e} Toulouse 3, INSA-T, Grenoble and Toulouse, France}

\author{Ermin Malic}
\affiliation{Department of Physics, Philipps-Universit{\"a}t Marburg, Renthof 7 35032 Marburg, Germany}

\author{Paulina P{\l}ochocka}\email{paulina.plochocka@lncmi.cnrs.fr}
\affiliation{Department of Experimental Physics, Faculty of Fundamental Problems of Technology, Wroclaw University of Science and Technology, 50-370 Wroclaw, Poland}
\affiliation{Laboratoire National des Champs Magn\'etiques Intenses, EMFL, CNRS UPR 3228, Universit{\'e} Grenoble Alpes, Universit{\'e} Toulouse, Universit{\'e} Toulouse 3, INSA-T, Grenoble and Toulouse, France}

\date{\today}

\maketitle

%\section{Electric field}

\section{Supporting measurements}

\begin{figure*}[ht]
\centering
\includegraphics[width=1.0\linewidth]{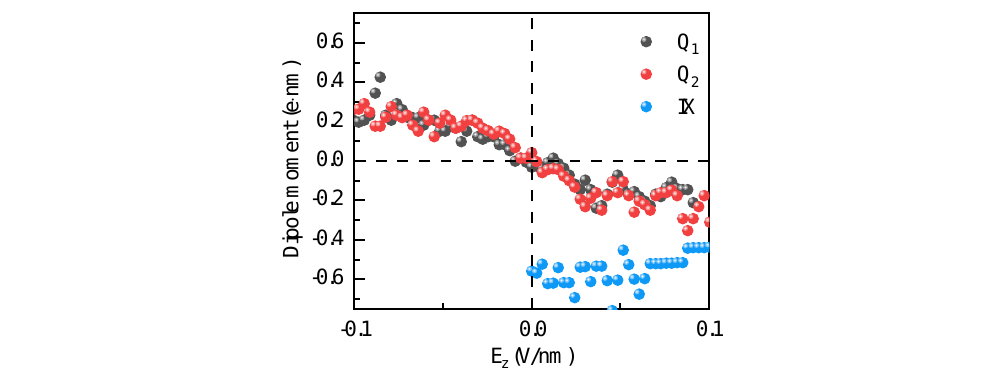}
\caption{\textbf{Electric field dependence of the static electric dipole moment.} Q$_1$ and Q$_2$ correspond to quadrupolar excitons and IX to the redshifting branch of spin-triplet interlayer exciton IX$^{\text{A}}_{1\text{s}\uparrow\downarrow}$.}
\label{fig:DipoleVsEz}
\end{figure*}

Fig.\ \ref{fig:DipoleVsEz} shows the electric field dependence of the static electric dipole moment of the Q$_1$, Q$_2$ and the IX$^{\text{A}}_{1\text{s}\uparrow\downarrow}$ (IX), calculated as $\text{d}E/\text{d}E_z$. The non-linear dependence of the quadrupolar exciton energy on the electric field translates to a vanishingly small electric dipole at low fields. The static dipole moment increases with increasing electric field and steadily approaches the dipole moment obtained for the dipolar spin-triplet interlayer exciton IX$^{\text{A}}_{1\text{s}\uparrow\downarrow}$.However, around 0.1 V/nm, the quadrupolar state begins to interact with other excitonic states due to the avoided crossing with A$_{1\text{s}}$ excitons, causing the dipole moments of the quadrupolar states to deviate from predictions based on a simple two-level model.

\clearpage

\begin{figure*}[th]
\centering
\includegraphics[width=1.0\linewidth]{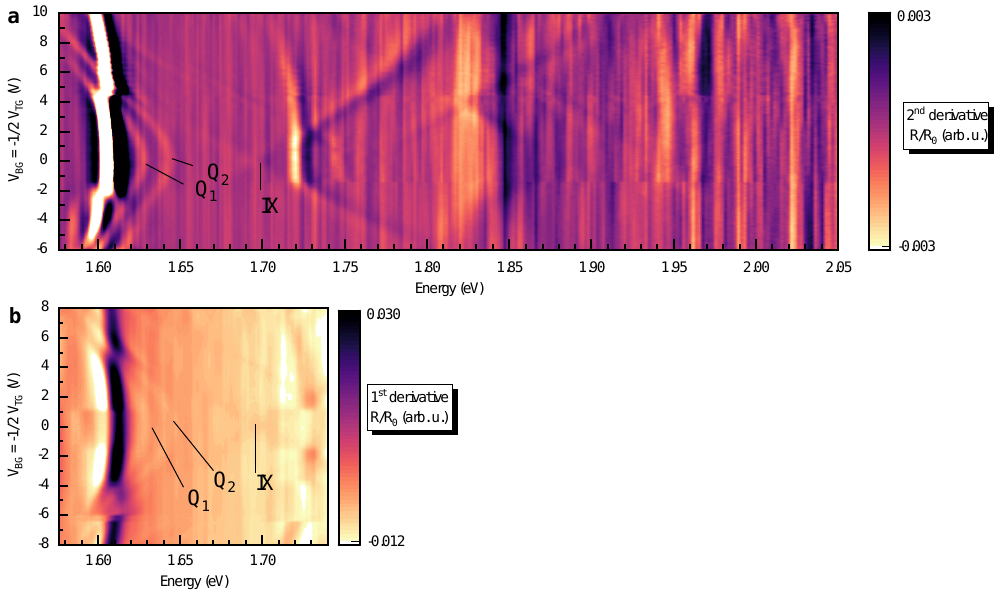}
\caption{\textbf{Reflectivity as a function of the electric field measured at different spots.} \textbf{a} and \textbf{b} False-color maps of two additional series of reflectivity spectrum measured as a function of $E_z$ on the main device. Reflectivity shown in the form of 2$^{\text{nd}}$ derivative $R/R_0$ in \textbf{a} and 1$^{\text{st}}$ derivative $R/R_0$ in \textbf{b}. The Q$_1$ and Q$_2$ branches are marked, as well as the above dipolar interlayer exciton IX.
}
\label{fig:additional_series}
\end{figure*}

\begin{figure*}[t]
\centering
\includegraphics[width=1.0\linewidth]{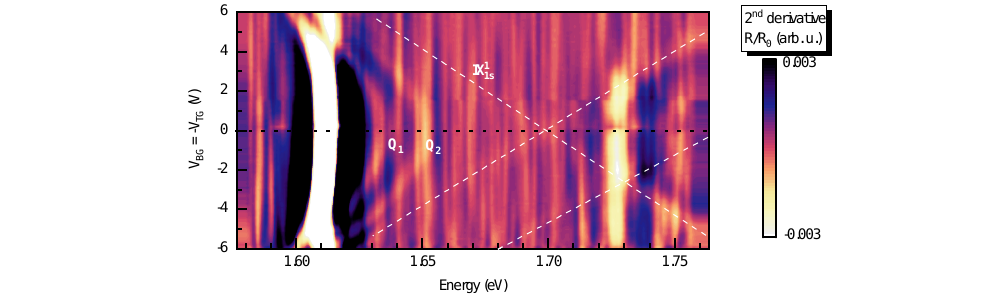}
\caption{\textbf{False-color map of 2$^{nd}$ derivative of reflectivity as a function of the out of plane electric field on the second natural MoSe$_2$ bilayer device.} Two quadrupolar branches Q$_1$ and Q$_2$ as well as the linearly shifting triplet IX states can be observed, similar as on the device shown in main text.
}
\label{fig:2nd}
\end{figure*}

We show in Fig.\,\ref{fig:additional_series} additional reflectivity spectra as a function of the electric field $E_z$ at two different spots (panels (a) and (b)). In all the investigated spots, we find evidence of the presence of quadrupolar excitons, as highlighted in Fig.\,\ref{fig:additional_series}. The consistent observation of these features in multiple spots on one sample, as well as their presence on another sample (see Fig.\ \ref{fig:2nd}), strongly suggests that the quadrupolar exciton complexes are an inherent feature of natural MoSe$_2$ bilayers.

Fig.\ \ref{fig:2nd} shows false-color map of 2$^{nd}$ derivative of reflectivity $R/R_0$ as a function of electric field (opposite voltage polarity applied to bottom and top gates) measured on a second natural MoSe$_2$ bilayer device. Similar to the device shown in the main text, here also two quadratically shifting quadrupolar branches Q$_1$ and Q$_2$, as well as linearly shifting spin-triplet IXs can be observed.

\clearpage

\begin{figure*}[t]
\centering
\includegraphics[width=1.0\linewidth]{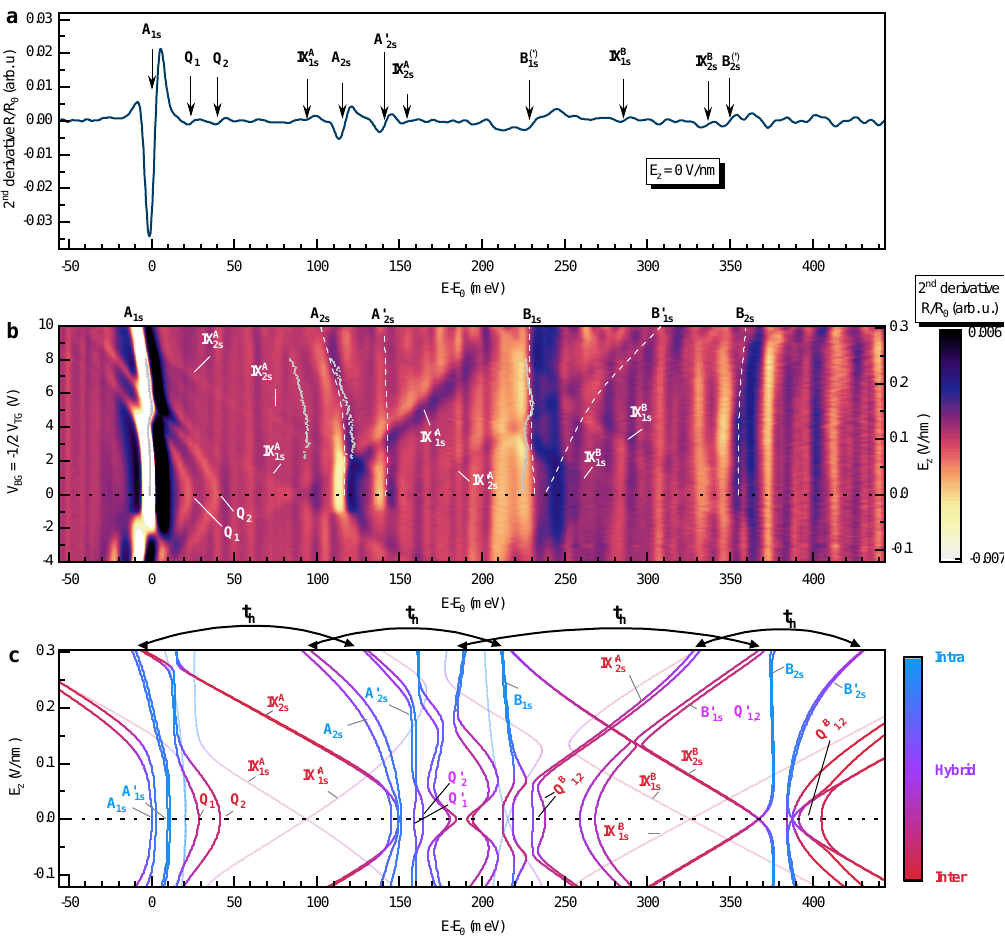}
\caption{\textbf{Measured and simulated exciton energy landscape under electric field.} \textbf{a} 2$^{\text{nd}}$ derivative of reflectivity at $E_z=\SI{0}{\volt / \nano\metre}$. \textbf{b} False-color map of 2$^{\text{nd}}$ derivative of reflectivity as a function of the out of plane electric field. The overlaid grey points (A$_{1s}$, A$_{2s}$, B$_{1s}$) are taken from the fitting of the PL spectrum shown in Fig.\ \ref{fig:PL}. In panels \textbf{a} and \textbf{b} the energy is measured relative to the A$_\textrm{1s}$ exciton at $E_0=1.606$\,eV. \textbf{c} Simulation of the excitonic landscape as a function of the out of plane electric field. The opaque and semi-transparent lines correspond to singlet and triplet states, respectively. The additional superscripts ``A'' and ``B'' denote the lower A-like and higher B-like interlayer states, respectively, exemplified in Fig.\ \ref{fig:states}(d).}
\label{fig:quad}
\end{figure*}

Fig.\,\ref{fig:quad} shows the evolution of the excitonic energy landscape in a broad spectral and electric field ($E_z$) range. Panel (a) shows the reflectivity spectrum at zero electric field, where various resonances related to intra- and interlayer transitions are marked.
Panel (b) shows the 2$^{nd}$ derivative of reflectivity $R/R_0$ as a function of the applied $E_z$. Aside from the quadratically shifting quadrupolar Q$_1$ and Q$_2$ excitons, several interesting features can be noted: (i) The avoided crossing and transfer of the oscillator strength between intralayer and interlayer transitions related to the hole tunneling, as in the case of the very strong mixing of the IX$^{\prime A}_{1s \uparrow\downarrow}$ and B$^\prime_{1s}$. (ii) The electron tunneling related anti-crossing between the A$_{1s}$ excitons and the Q$_1$, Q$_2$ and IX$^{A}_{1s \uparrow\downarrow}$ at $E_z \sim$ 0.1 -- 0.15 V/nm as well as with higher lying 2s IXs at higher $E_z$. (iii) At $E_z \sim$ 0.1 -- 0.15 V/nm we observe transfer of the oscillator strength between the spin-singlet IX$^{A}_{2s \uparrow\uparrow}$ and lower energy state (detuned by approximately 20 meV) likely related to brightened spin-triplet IX$^{\prime A}_{2s \uparrow\downarrow}$.
Panel (c) shows the calculated excitonic landscape under electric field, which includes all three considered couplings. Here we can observe the very good qualitative agreement with the experimental results shown in panel (b). Particularly, the emergence of the anti-crossing at the A$_{1s}$ exciton corroborating the experimental findings as well as several examples of hole tunneling related hybridizations (marked by purple arrows), which are also visible in the experiment.
The aforementioned 2s spin-triplet state IX$^{\prime A}_{2s \uparrow\downarrow}$, the emergence of which can be observed in the reflectivity at high $E_z$, does not appear in the simulated spectra in panel (c) because these 2s spin-triplet states are not considered in the model. This is because they are not essential to understanding the formation of quadrupolar excitons, which are the focus of this study.

\clearpage

\begin{figure*}[t]
\centering
\includegraphics[width=1.0\linewidth]{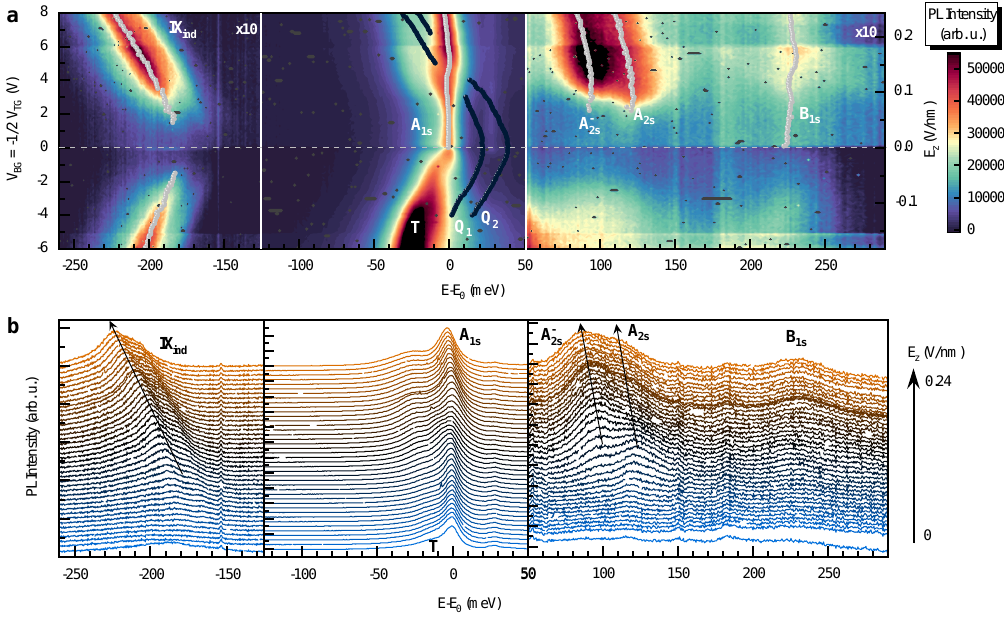}
\caption{\textbf{Photoluminescence responce under the electric field.} \textbf{a} False-color map of PL as a function of the out of plane electric field $E_z$. The overlaid black data points (Q$_1$ and Q$_2$) correspond to the energy position extracted from the Reflectivity dataset shown in Fig.\ \ref{fig:quad}. The overlaid grey points (A$_{1s}$, A$_{2s}$, B$_{1s}$) are taken from the fitting of the PL emission shown here. The quadrupolar transitions  Q$_1$ and Q$_2$ are taken as the minima in the reflectivity spectrum shown in Fig.\ \ref{fig:quad}(b). We note here that the application of the gate voltages is not symmetrical and for the negative $E_z$ field direction the sample shows significant doping, thus we focus on the positive $E_z$ part. \textbf{b} Waterfall plot of the PL spectrum from $E_z$=0 V/nm to $E_z$ = 0.24 V/nm. Spectra are shifted vertically for clarity. The energy scale (E-E$_0$) is shown relative to the A$_\textrm{1s}$ exciton at $E_0=1.606$\,eV.
}
\label{fig:PL}
\end{figure*}

Fig.\ \ref{fig:PL} shows the photoluminescence (PL) spectrum as a function of the applied electric field in the form of a false-color map (panel (a)) and stacked individual spectra (panel (b)). At $E_z = \SI{0}{\volt/\nano\metre}$, the strongest PL peak originates from the A$_{1s}$ transition. On the low energy side of A$_{1s}$, the less intense negatively trion (T) emission is observed. At low energy side ($\sim -\SI{170}{\milli\eV}$ below A$_{1s}$) a weak, momentum-indirect interlayer exciton (IX$_{ind}$) can be observed. For increasing $E_z$ (positive direction) we observe linear red shift of the aforementioned IX$_{ind}$, the slope of which is approximately half of the one characterizing the momentum-direct IXs visible in the reflectivity map of Fig.\ \ref{fig:quad}(b). This is due to the IX$_{ind}$, which stems from the $K-\Gamma$ transition \cite{sung2020broken,kovalchuk2023interlayer,villafane2023twist}, where the hole at $\Gamma$ point is delocalized between the two layers, effectively lowering the dipole moment. Moreover, increasing $E_z$ results in the enhancement of the PL intensity of the A$_{2s}$ and B$_{1s}$ states. Interestingly, a lower energy side peak to the main A$_{2s}$ peak appears, which most likely originates from the negatively charged A$^-_{2s}$ trion, similar to what was observed in monolayer WSe$_2$ \cite{sell2022magneto}. At the same time, the low energy side peak of A$_{1s}$, attributed to trion (T) at $E_z = \SI{0}{\volt/\nano\metre}$ also gains intensity in the same $E_z$ range. The presence of charged exciton features in the PL spectrum points to a small degree of unintentional doping, as the reflectivity lacks these signatures. Furthermore, the PL peak energies shift rapidly, due to anti-crossings with otherwise non-emissive interlayer dipolar or quadrupolar excitons. This behaviour is visible at $E_z\sim$ 0.1 -- 0.15 V/nm for both A$_{1s}$ and B$_{1s}$ excitons, which corresponds to the electric field range where avoided crossings for these excitonic species are observed in the reflectivity spectrum of Fig.  \ref{fig:quad}(b), where PL peak energies are also overlaid (grey points).

\clearpage

\begin{figure*}[t]
\centering
\includegraphics[width=1.0\linewidth]{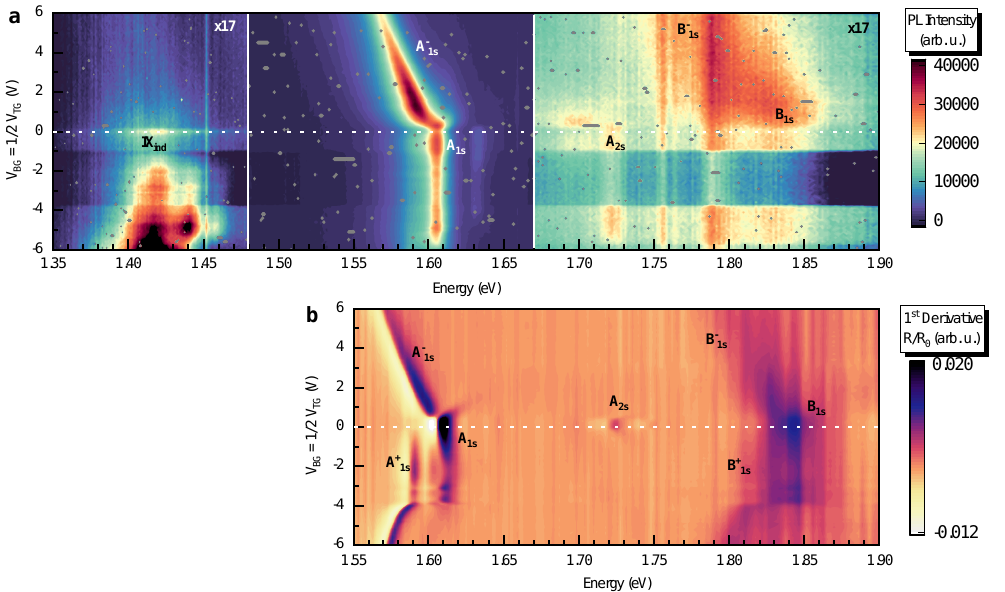}
\caption{\textbf{Free carrier doping dependent photoluminescence and reflectivity.} False-color map of \textbf{a} the PL spectrum and of the \textbf{b} 1$^{st}$ derivative of reflectivity as a function of the gate voltages (with constant ratio $V_{BG}=\frac{1}{2}V_{TG}$) corresponding to the change in free carrier concentration. Both PL and reflectivity spectra show a clear switching between different doping regimes. At $V_{BG}=\frac{1}{2}V_{TG}=0 V$, we see the most neutral regime due to the highest absorption strength of the neutral complexes of A$_{1s}$, A$_{2s}$ and B$_{1s}$ with minimized contribution of charged complexes. Thus those voltages were chosen as an optimal starting point for the electric field sweep.
}
\label{fig:Doping}
\end{figure*}

Fig.\ \ref{fig:Doping} shows the PL spectrum (panel (a)) and 1$^{st}$ derivative of the reflectivity $R/R_0$ (panel (b)) as a function of the gate voltages applied with the same polarity, which enables free carrier doping, without simultaneous application of electric field. The gate voltage ratio accounting for the unequal thickness of bottom and top hBNs was chosen the same (in absolute value) as for the electric field sweep, i.e., V$_{BG}=\frac{1}{2} V_{TG}$. As can be observed in the shown figures, the minimal carrier doping, as observed by smallest contribution of charged excitonic states (attractive Fermi-polarons \cite{sidler2017fermi}) as well as most prominent presence of neutral excitonic states (repulsive Fermi-polarons \cite{sidler2017fermi}), is in the vicinity of V$_{BG}=\frac{1}{2} V_{TG} = 0$ V. For positive gate voltages, direction we can observe the change of the regime into electron doped regime, as attested by the emergence of the negatively charged A$_{1s}$ exciton  feature, which shifts away from the neutral exciton due to increasing doping level. Consequently, the negative gate voltages polarity changes the doping regime to the hole doping, as seen by the emergence of positively charged exciton feature.

\clearpage

%-----------------------------------
\section{Theoretical model}
%-------------------------------------
In order to form quadrupole excitons we need a coupling between differently
oriented dipoles. In its simplest form such a Hamiltonian can be written as
%simple starting Hamiltonian (main)
\begin{equation} \label{eq1}
H =  \begin{pmatrix}
E_{IX}+d_LE_z & J \\
J^* & E_{IX'}-d_LE_z,
\end{pmatrix} 
\end{equation}

where $E_{IX}$ is the interlayer exciton energy of one dipole and $E_{IX'}$ is the energy
of the reversed dipole. The linear energetic shift with the electrical field is
described by $d_L$$E_z$, where $d_L$ is the dipole length and $E_z$ is the field strength.
$J$ is the coupling between the interlayer excitons. Diagonalizing
such a Hamiltonian yields two quadrupole excitons with opposite curvature which we call Q and Q', and they will be energetically separated by the energy $2J$.

Starting with the single particle picture, terms that could form quadrupole excitons will emerge from the Coulomb interaction,

%Generic Coulomb interaction Hamiltonian
\begin{equation} \label{eq2}
H_{C}=\frac{1}{2}\sum_{1234}^{}\bar{V}^{1234}a_{1}^{\dagger}a_{2}^{\dagger}a_{3}a_{4},
\end{equation}
where 1 − 4 are generic compound indices and $\bar{V}^{1234}$ is the Coulomb matrix
element. Assuming low densities, the form of the Coulomb Hamiltonian that
will mix different dipoles will have the following form,

%Specific Coulomb interaction Hamiltonian
\begin{equation} \label{eq3}
H_{C}=\sum_{ss'\xi_{\lambda}\xi'_{\lambda}l_{\lambda}l'_{\lambda}\textbf{k}\textbf{k'}\textbf{q}}^{}\bar{V}_{ss'\xi_{\lambda}\xi'_{\lambda}l_{\lambda}l'_{\lambda}}^{CV} (\textbf{k},\textbf{k'},\textbf{q})v^{\dagger}_{\xi'_hs'l'_h,\textbf{k'}-\textbf{q}}c^{\dagger}_{\xi_esl_e,\textbf{k}+\textbf{q}}c_{\xi'_es'l'_e,\textbf{k'}}v_{\xi_hsl_h,\textbf{k}}
\end{equation}
where $s^{(')} = (\uparrow, \downarrow)$ is the spin index, $\xi^{(')}_\lambda$ is the valley index and $l^{(')}_\lambda$ is the layer
index with $\lambda$ = (c, v). The relative momenta is given by k and k' respectively,
and the transferred momenta is given by q. Furthermore, c$^{(\dagger)}$/$v^{(\dagger)}$ are annihilation(creation) operators for the conduction band/valance band. Switching
to the exciton basis, and only keeping the relevant mixing terms gives,

%Transferring to exciton basis
\begin{equation} \label{eq4}
\begin{split}
H_{C}=\sum_{\substack{\tilde{s}\tilde{s}'\xi\xi'LL'\\\textbf{k}\textbf{k'}\textbf{q}}}^{}\bar{V}_{\tilde{s}\tilde{s}',\xi\xi'LL'}^{CV} (\textbf{k},\textbf{k'},\textbf{q})X^{\dagger}_{\xi\tilde{s}L,\textbf{q}}X_{\xi'\tilde{s}'L',\textbf{q}}\times \phi^{*\xi\tilde{s}l_e}_{\xi\tilde{s}l_h}(\textbf{k}+\alpha^\xi\textbf{q})\phi^{\xi'\tilde{s}'l'_e}_{\xi'\tilde{s}'l'_h}(\textbf{k'}-\beta^{\xi'}\textbf{q}).
\end{split}
\end{equation}

Here L = (l$_e$, l$_h$) is a compound layer index,s = $(\uparrow\uparrow, \downarrow\downarrow, \downarrow\uparrow, \uparrow\downarrow)$ is a compound
spin index (the spin triplet states will not be spin conserving and the matrix
element will therefore be set to 0 for these). Note that in electron-hole picture,
the spin for the hole would be reversed. The compound valley index is given by $\xi = (\xi_e, \xi_h)$ and X$^{(\dagger)}$ are exciton annihilation(creation) operators. Simplifying the above expression to only include the relevant terms for the quadropole coupling gives us

%tilde J coupling hamiltonian
\begin{equation} \label{eq5}
H_{QC}=\sum_{\tilde{s}\tilde{s}'\xi\xi'LL'\textbf{Q}}^{}
\tilde{J}_{\tilde{s}'\xi'L'}^{\tilde{s}\xi L} (\textbf{Q})
X^{\xi\tilde{s}\dagger}_{L\textbf{Q}}
X^{\xi'\tilde{s}'}_{L',\textbf{Q}}, 
\end{equation}

where we have changed the matrix element to $\tilde{J}_{\tilde{s}'\xi'L'}^{\tilde{s}\xi L} (\textbf{Q})$, which only includes the specific contributions of the Coulomb Hamiltonian that mixes different dipoles and the exciton wave functions. Moreover, we have changed notation of the transferred momentum so it directly
corresponds to the center-of-mass momentum \textbf{Q}. The matrix element $\tilde{J}_{\tilde{s}'\xi'L'}^{\tilde{s}\xi L} (\textbf{Q})$ can thus be written out as

\begin{equation}
    \tilde{J}_{\tilde{s}'\xi'L'}^{\tilde{s}\xi L} (\textbf{Q})=\sum_{\bm{k}^{\prime}\bm{k}}\bar{V}_{\tilde{s}\tilde{s}',\xi\xi'LL'}^{CV} (\textbf{k},\textbf{k'},\textbf{Q})\phi^{*\xi\tilde{s}l_e}_{\xi\tilde{s}l_h}(\textbf{k}+\alpha^\xi\textbf{Q})\phi^{\xi'\tilde{s}'l'_e}_{\xi'\tilde{s}'l'_h}(\textbf{k'}-\beta^{\xi'}\textbf{Q}).
\end{equation}

The Coulomb matrix element can in turn be derived as \cite{yu2014dirac}

\begin{equation}\label{eq:ColMatrix}
    \begin{split}
        \bar{V}_{\tilde{s}\tilde{s}',\xi\xi'LL'}^{CV} (\textbf{k},\textbf{k'},\textbf{Q})=\sum_{\bm{G}}\frac{V(\bm{G}+\bm{Q})}{A}\Big\langle\mathcal{U}^{cl_e}_{\xi,\bm{k}+\bm{q},s}(\bm{r})\Big|e^{i\bm{G}\cdot \bm{r}_1}\Big|\mathcal{U}^{vl_h}_{\xi,\bm{k},s}(\bm{r})\Big\rangle\\
        \times\Big\langle\mathcal{U}^{vl_h^{\prime}}_{\xi^{\prime},\bm{k}^{\prime}-\bm{q},s^{\prime}}(\bm{r})\Big|e^{-i\bm{G}\cdot \bm{r}_1}\Big|\mathcal{U}^{cl_e^{\prime}}_{\xi^{\prime},\bm{k}^{\prime},s^{\prime}}(\bm{r})\Big\rangle,
    \end{split}
\end{equation}

where $\mathcal{U}^{\lambda l_{\lambda}}_{\xi,\bm{k},s}(\bm{r})$ are the Bloch factors to the electronic Bloch wave function $\Psi^{\lambda l_{\lambda}}_{\xi,\bm{k},s}(\bm{r})=e^{i(\bm{\xi}+\bm{k})\cdot \bm{r}}\mathcal{U}^{\lambda l_{\lambda}}_{\xi,\bm{k},s}(\bm{r})$ and $\bm{G}$ are the reciprocal vectors of the lattice.

\begin{figure*}[h]
\centering
\includegraphics[width=1.0\linewidth]{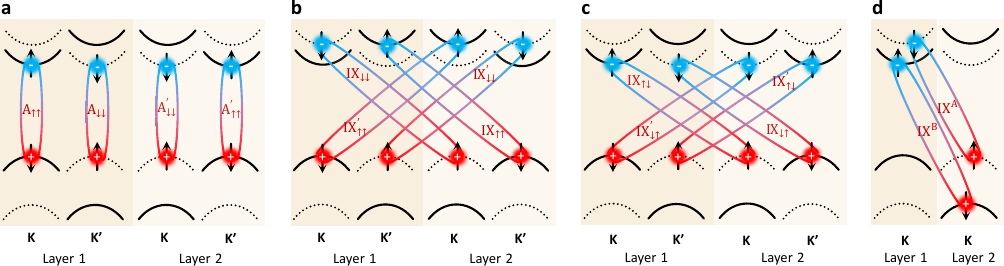}
\caption{\textbf{Excitonic states in MoSe$_2$ homobilayer.} Schematic showing the four degenerate states of A exciton \textbf{a}, spin-singlet IX \textbf{b}, spin-triplet IX \textbf{c}. When labelling the exciton state, the first arrow in the subscript denotes the spin of the electron and the second of the hole (which is reversed in the electron-hole picture). The prime superscript denotes a reversed dipole moment, which corresponds to switched position (layer) of the electron and the hole. \textbf{d} Example of the lower(higher) lying A-like (B-like) spin-singlet IX denoted with superscript ``$^{\text{A(B)}}$'', with the hole residing in higher(lower) valence band.
}
\label{fig:states}
\end{figure*} 

As can be seen from Fig. \ref{fig:states} we have two distinct types of mixing between different dipoles. One case with the same valley, but different spin and one case with different spin, but the same valley. Consequently, we can rewrite the Hamiltonian into these two parts

%J coupling hamiltonian with J' and J
\begin{equation} \label{eq6}
H_{QC}=\sum_{\tilde{s}\tilde{s}'\xi\xi'LL'\textbf{Q}}^{}
\Big(J_{\tilde{s}'\xi L'}^{\tilde{s}\xi L} (\textbf{Q})\delta_{\xi\xi'}+J_{\tilde{s}\xi'L'}^{\prime\tilde{s}\xi L} (\textbf{Q})\delta_{\tilde{s}\tilde{s}'}\Big)
X^{\xi\tilde{s}\dagger}_{L\textbf{Q}}
X^{\xi'\tilde{s}'}_{L',\textbf{Q}}.
\end{equation}

Here, $J$ and $J^{\prime}$ are the matrix elements for the two different forms of dipole exchange processes taken into consideration. For $J^{\prime}$, we have a mixing of different valleys, which consequently means we only have long-range interaction \cite{yu2014dirac}. This corresponds to $\bm{G}=0$ in equation\,S\ref{eq:ColMatrix} and would be proportional to the optical dipole matrix element. The scenario is different for $J$ were $\bm{G\neq}0$ is also allowed due to the valleys being the same, this short-range addition to the term would then make $J>J^{\prime}$. Matching these to experiments reveal that $J=90$ meV and $J^{\prime}=8$ meV. These are much larger than the expected values, which should lie close to the value of the optical dipole matrix element, which for interlayer excitons are orders of magnitude smaller. 

In our considered material, the situation becomes more complex than just
two different interlayer excitons. In naturally stacked MoSe$_2$ we have four degenerate
excitons. Two from layer degeneracy (K and K') and two from spin degeneracy,
which can be seen for the four different A excitons (Fig. S1a). Due to one layer
being stacked 180$^\circ$ with respect to the other (H-type stacking), the spin-orbit coupling can be
considered as reversed in one layer. Therefore, the four degenerate bright A excitons
stem from a combination of different valleys (K or K'), different spin and
different layer. The same scenario happens for the bright spin-allowed interlayer
exciton states (Fig. S1b). We have four degenerate interlayer excitons with
different spin, valley configurations and layer index. Two of them will have the
reversed dipole moment.
The interlayer excitons and intralayer excitons can now couple to each other
via electron or hole tunneling. Normally, electron tunneling is symmetry forbidden
in pure naturally stacked homobilayers \cite{PhysRevB.95.115429}. The visible avoided crossing
between the quadrupole excitons and the A exciton seen in the experiment
suggests however that some limited electron tunneling is allowed. The carrier
tunneling can be modeled in accordance with ref \cite{PhysRevResearch.3.043217}. By including the new dipole exchange
coupling we can now write down an extended Hamiltonian for field dependent exciton energy landscape

%Simplified full Hamiltonian (should go to main)
\begin{equation}\label{eq7}
\begin{split}
H & =H_{0}+H_T+H_{QC},
\end{split}
\end{equation}

where $H_0$ includes the bare exciton binding energies as obtained from the generalized Wannier equation \cite{ovesen2019interlayer} and the exciton response to the external electric field. The tunneling Hamiltonian is given by $H_T$, which takes into account both electron and hole tunneling. The last contribution is the new dipole exchange coupling giving rise to the formation of quadropole excitons. Writing out the full Hamiltonian gives us

%Last full Hamiltonian
\begin{equation}\label{eq8}
\begin{split}
H & =\sum_{\xi\xi'\tilde{s}\tilde{s}'\textbf{Q}LL'\mu\nu^{}}\Bigg( \Big(E_{L\textbf{Q}}^{\tilde{s}\xi\mu}+d_{L}E_{z}\Big)X_{L\textbf{Q}}^{\mu \tilde{s}\xi\dagger}X_{L\textbf{Q}}^{\mu \tilde{s}\xi}  \delta_{LL'}\delta_{\tilde{s}\tilde{s}'}\delta_{\xi\xi'}\delta_{\mu\nu}  
\\& + T_{LL'}^{\mu\nu\xi} X_{L\textbf{Q}}^{\mu\tilde{s}\xi\dagger} X_{L'\textbf{Q}}^{\nu\tilde{s}\xi} \delta_{\tilde{s}\tilde{s}'} \delta_{\xi\xi'}  + \Big(J_{\tilde{s}'\xi L'}^{\tilde{s}\xi L} (\textbf{Q})\delta_{\xi\xi'}+J_{\tilde{s}\xi'L'}^{\prime\tilde{s}\xi L} (\textbf{Q})\delta_{\tilde{s}\tilde{s}'}\Big)
X^{\xi\tilde{s}\dagger}_{L\textbf{Q}}
X^{\xi'\tilde{s}'}_{L',\textbf{Q}}\Bigg),
\end{split}
\end{equation}

where $E_{L\textbf{Q}}^{\tilde{s}\xi\mu}$ is the bare exciton dispersion and $d_LE_z$ is the induced shift from the electrical field. Moreover, $T_{LL'}^{\mu\nu\xi}$ is the tunneling matrix element. Here, $\mu(\nu)$ are the exciton state indices summing over 1s and 2s. The dipole exchange coupling is given by $J$ and $J'$, where we have assumed that the coupling between 1s and 2s in J is small, thus only taking into account 1s for $J/J'$.

\begin{figure*}[t]
\centering
\includegraphics[width=1.0\linewidth]{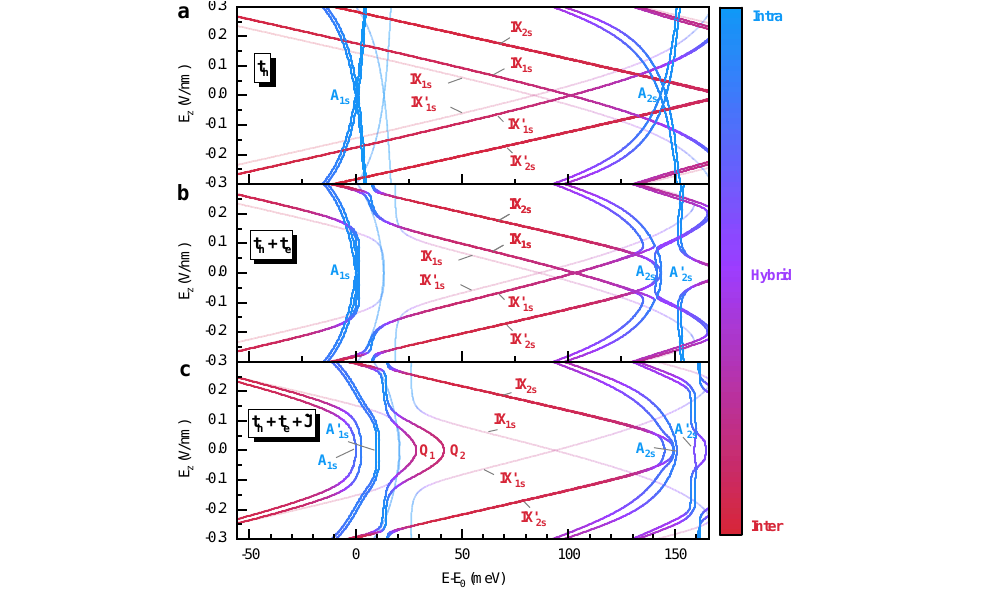}
\caption{\textbf{Simulated exciton energy landscape under the electric field with different coupling contributions.} Simulated evolution of the excitonic landscape in presence of the out of plane electric fields by inclusion of \textbf{a} hole tunneling, \textbf{b} hole and electron tunneling, \textbf{c} hole and electron tunneling as well as the new coupling dipole exchange coupling $\tilde{J}=J+J'$ . The opaque and semi-transparent lines correspond to singlet and triplet states, respectively. The energy scale (E-E$_0$) is shown relative to the A$_{1\text{s}}$ exciton energy $E_0$.
}
\label{fig:couplings}
\end{figure*}

The complete Hamiltonian now gives rise to several different coupling channels.
As an example the IX$_{\uparrow\uparrow}$ exciton will couple to IX'$_{\downarrow\downarrow}$ via the new coupling $\tilde{J}$, but each of the interlayer excitons will also couple to their respective A exciton via electron tunneling. Furthermore, they will also couple to
the higher lying B exciton via hole tunneling. For the purpose of this effective
model we have only considered a coupling between the four degenerate interlayer
excitons in the coupling $J/J'$. Diagonalizing the Hamiltonian now gives us the final exciton energies.

In Fig. \ref{fig:couplings}, the calculated exciton energies are shown as a function of electric field by including one coupling at a time. First, Fig. \ref{fig:couplings}(a) shows the case in which only hole tunneling is considered.
Here, we can see multiple different shifts from the various interlayer excitons. Additionally the strong mixing of both A$_{1s}$ and A$_{2s}$ excitons with higher lying interlayer excitons due to hole tunneling results in their energy shift (splitting at higher electric fields) and change of the intra-inter layer character (See Fig. \ref{fig:quad}(c)) where all hole tunneling related hybridizations are marked).
By then including electron tunneling (Fig. \ref{fig:couplings}(b)) we can see the
emergence of avoided crossings with the A exciton when the interlayer excitons
pass through. Since the electron tunneling is much less efficient than the hole tunneling, the hybridization is far more localized around the crossing. Finally,
by including the new couplings $J/J'$ (Fig. \ref{fig:couplings}(c)) we can see a clear emergence of quadrupolar excitons Q$_1$ and Q$_2$, exhibiting their characteristic parabolic
shape. The higher lying quadrupoles are nearly completely suppressed due to
tunneling with the 2s state, where we as a consequence can see a splitting of
the 2s state at 0 electrical field. We can also note that the 2s state will split
at higher electrical field strengths due to efficient hybridization with the higher
lying 2s interlayer exciton.

The splitting of the A exciton stems from a combination of hole tunneling
and $J/J'$. Due to the hole tunneling with higher lying interlayer excitons, it will
exhibit a very small, but non-zero interlayer character, which in turn means that
it will be affected by the coupling J (same valleys, but different spin) and split some of the A excitons in an
attempt to form quadrupoles. Similarly, a very small effect from J' (different valleys, same spins) can also be
seen in the two lowest branches of the A exciton. Due to the weak hybridization
of the A exciton, both effects are very much suppressed however. Note here,
that all four A excitons are still mostly degenerate. The four
different branches of the A exciton all exhibits a superposition of the original
A exciton constituents and the splitting is only due to their
very small interlayer nature. At larger electrical field strengths we can see
the quadrupoles Q$_1$ and Q$_2$ cross the A exciton branches and via electron
tunneling forming avoided crossings and then continuing to become the lower
lying branches themselves. Here, we have also taken into account the spin triplet
exciton states IX$_{\uparrow\downarrow}$ and IX$'_{\downarrow\uparrow}$. Since a coupling between these excitons would not be spin conserving it should not experience the coupling J. Therefore, the triplet interlayer exciton states follow the standard linear shift with electrical field instead.

\clearpage

%-------------------------------------------------
\section{Analysis of the reflectivity spectra}
%----------------------------------------------------
We show here briefly how the reflectivity spectra have been analyzed to obtain the false-color maps shown in the main article and in the Supplementary Information. In Fig.\ \ref{fig:showcase}(a), we show a representative reflectivity spectrum $R$ measured on the MoSe$_2$ bilayer and the reference spectrum $R_0$ used to normalize all the spectra, measured on the SiO$_2$ substrate. The choice of acquiring the reference spectrum on the SiO$_2$ substrate was dictated by the non-uniform hBN and graphene thickness in areas which did not include any TMD layer. Starting from the raw $R$ and $R_0$, spectra their ratio was initially calculated, as shown in Fig.\ \ref{fig:showcase}(b). After the first derivative with respect to the wavelength was computed and smoothed, see Fig.\ \ref{fig:showcase}(c), a background subtraction, common for all the spectra measured as a function of the electric field in a specific spot, was performed, as illustrated in Fig.\ \ref{fig:showcase}(d). Subsequently, the second derivative spectrum was calculated (derivation with respect to the wavelength), with the smoothing parameters specified in Fig.\ \ref{fig:showcase}(e).
\begin{figure*}[t]
\centering
\includegraphics[width=1.0\linewidth]{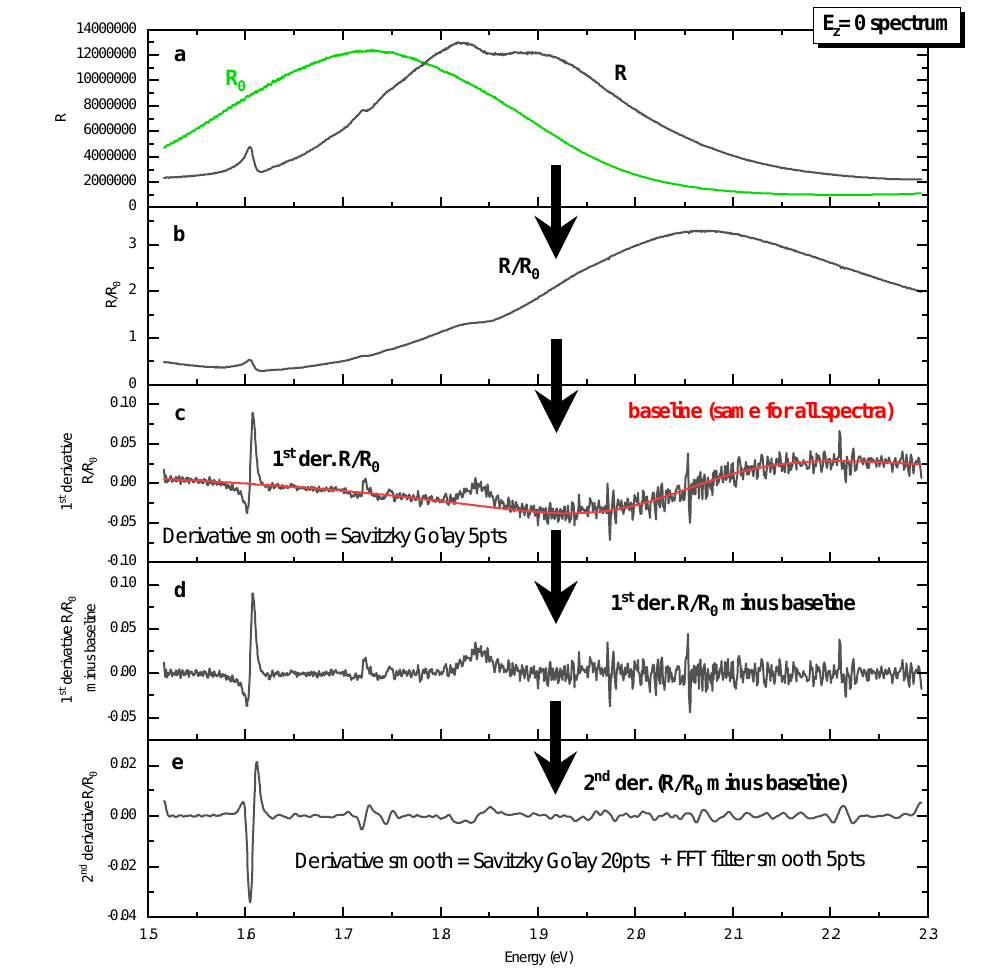}
\caption{\textbf{Analysis of the reflectivity spectra.} \textbf{a} Raw reflectivity spectrum $R$ measured on the MoSe$_2$ bilayer and reference spectrum $R_0$ measured on the SiO$_2$ substrate. \textbf{b} Reflectivity ratio $R/R_0$ and \textbf{c} first derivative with respect to the wavelength. In red, the baseline of the derivative spectrum is shown. The smoothing parameters are indicated. \textbf{d} First derivative after the baseline subtraction. \textbf{e} Second derivative with the corresponding smoothing parameters.}
\label{fig:showcase}
\end{figure*} 

\begin{figure*}[ht]
\centering
\includegraphics[width=1.0\linewidth]{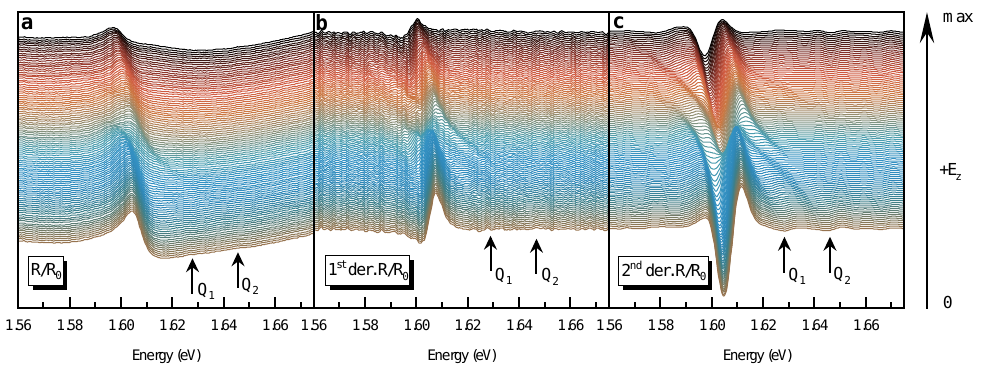}
\caption{\textbf{Reflectivity spectra under the electric field.} Vertically stacked reflectivity spectra as a function of $E_z$ (positive direction) in the form of \textbf{a} $R/R_0$ \textbf{b} 1$^{\text{st}}$ derivative $R/R_0$ and \textbf{c} 2$^{\text{nd}}$ derivative $R/R_0$. The Q$_1$ and Q$_2$ transitions are marked on the zero-field spectra.}
\label{fig:raw_spectra}
\end{figure*}

In Fig.\,\ref{fig:raw_spectra} we show the comparison of the reflectivity spectra as a function of the electric field (positive direction) in the form of $R/R_0$, 1$^{\text{st}}$ derivative $R/R_0$ and 2$^{\text{nd}}$ derivative $R/R_0$ in panels\,(a,b,c), respectively. The quadratically shifting quadrupolar states Q$_1$ and Q$_2$ are only weakly visible on the raw $R/R_0$ spectra (panel\,(a)). Calculating the 1$^{\text{st}}$ (panel\,(b)) and then 2$^{\text{nd}}$ (panel\,(c)) derivative of the $R/R_0$ spectra substantially improves the visibility of these weak spectral features.

\clearpage

\bibliography{Bibliography}